\documentclass[onecolumn,superscriptaddress,amsmath,amssymb,floatfix]{revtex4}

\usepackage{graphicx}
\usepackage{dcolumn}
\usepackage{bm}
\usepackage{textcomp}
\usepackage{color}
\usepackage{setspace}
\usepackage{txfonts}
\begin{document}
\title{Formation of Kinneyia via shear-induced instabilities in microbial mats}
\author{Katherine Thomas}
\email{katherine.thomas@ds.mpg.de}
\affiliation{Max Planck Institute for Dynamics and Self-Organization, Am Fassberg 17, D-37077 G\"ottingen, Germany}
\author{Stephan Herminghaus}
\affiliation{Max Planck Institute for Dynamics and Self-Organization, Am Fassberg 17, D-37077 G\"ottingen, Germany}
\author{Hubertus Porada}
\affiliation{University G\"ottingen, Geowissenschaftliches Zentrum, Goldschmidtstr. 3, D-37077 G\"ottingen, Germany}
\author{Lucas Goehring}
\affiliation{Max Planck Institute for Dynamics and Self-Organization, Am Fassberg 17, D-37077 G\"ottingen, Germany}
\date{\today}

\begin{abstract}
Kinneyia are a class of microbially mediated sedimentary fossils. Characterised by clearly defined ripple structures, Kinneyia are generally found in areas that were formally littoral habitats and covered by microbial mats. To date there has been no conclusive explanation of the processes involved in the formation of these fossils. Microbial mats behave like viscoelastic fluids. We propose that the key mechanism involved in the formation of Kinneyia is a Kelvin-Helmholtz type instability induced in a viscoelastic film under flowing water. A ripple corrugation is spontaneously induced in the film and grows in amplitude over time. Theoretical predictions show that the ripple instability has a wavelength proportional to the thickness of the film. Experiments carried out using viscoelastic films confirm this prediction. The ripple pattern that forms has a wavelength roughly three times the thickness of the film. This behaviour is independent of the viscosity of the film and the flow conditions. Laboratory-analogue Kinneyia were formed via the sedimentation of glass beads, which preferentially deposit in the troughs of the ripples. Well-ordered patterns form, with both honeycomb-like and parallel ridges being observed, depending on the flow speed. These patterns correspond well with those found in Kinneyia, with similar morphologies, wavelengths and amplitudes being observed.

\end{abstract}

\maketitle

\section{Introduction}
Biofilms are ubiquitous. Found in environments ranging from Antarctic ice \cite{Friedmann1987,Thomas2002} to the walls of deep sea hydrothermal vents \cite{Wolff1977} and the human body \cite{HallStoodley2009}, biofilms represent one of the earliest life forms on Earth. Recent discoveries provide evidence that  fossils of microbial origin date back to 3,450 million years ago \cite{Allwood2006}, suggesting that biofilms have been present throughout the majority of the Earth's history.

Microbial mats can be seen as thick biofilms. They range in thickness from a few millimetres to a few centimetres, depending on the habitat, growth conditions and bacteria that form them \cite{Gerdes2000,Neu1997}. One of the most successful mat forming bacteria is cyanobacteria due to its large morphologic variability and capacity for biostabilisation (increasing sediment stability or reducing erosion)  \cite{Gerdes2007}. However, mat formation is not limited to cyanobacteria and can result from organisms including bacteria, archaea \cite{Zolghadr2010}, protozoans, algae \cite{Brouwer2005} and fungi \cite{Verstrepen2006}. The term `microbial mat' will be used here to describe any macroscopic cohesive microbial community growing on, adhering to, or enmeshing an inorganic substrate \cite{Krumbein2003}.

Modern microbial mats are commonly found growing on rock, soil and granular carbonatic and siliciclastic surfaces in water-based habitats and are preferentially observed to grow in the intertidal to lower supratidal zones of riverine and marine environments \cite{Gerdes2000,Gerdes2007} as well as hypersaline lagoons \cite{Allen2009}. Microbial mats can enhance stabilisation of sandy substrata via the secretion of adhesive mucilages, which glue the sediment grains together, reducing the erodability of the sediment \cite{Schieber2007, Gehling1999, Pfluger1999, Noffke2001, Paterson1994}. The prevalence of biofilm systems, in particular cyanobacteria, have made them successful at leaving traces in sediments. This is readily observed in the host sediments of modern microbial mats \cite{Gerdes2007}.

Microbially induced sedimentary structures (MISS) can be grouped depending on the role that the microbial mat had in the formation process \cite{Noffke2010} and arise from physical processes including biostabilisation \cite{Cuadrado2011}, baffling and trapping \cite{Black1933, Reid2000}, binding \cite{Noffke2008} and growth \cite{Aref1998,Gerdes2000}. The patterns and structures arise due to the presence of microbial mats. In the absence of a mat no structures are expected to form. MISS are highly diverse \cite{Noffke2011}, with features ranging from the millimetre to metre length scales. They include wrinkle structures, palimpsest ripples, roll up structures and laminar structures \cite{Noffke2001, Noffke1996}. Recently Bouougri and Porada \cite{Bouougri2012} documented various structures including crack patterns and crumpled layers, which form when strong winds blow over microbial mats. While chemical processes are not considered here,  recent studies on modern microbial mats have shown that microbially mediated mineral precipitation  is important for the preservation of sedimentary structures \cite{Cuadrado2012, Darroch2012}. In siliciclastic environments precipitation of hydrated alumino-silicates from the mats may also result in cementation and lithification of the surrounding sediments, potentially also playing a stabilizing role in fossil preservation \cite{Meyer2012, Anderson2011, Laflamme2011}.

\subsection{Fossilised microbial mats}

Structures similar to those observed in modern living mats are also observed in the ancient geologic record \cite{Bouougri2007, Gerdes2007, Noffke2010}. Sedimentary features in the clastic sedimentary rock records from the Precambrian era point towards the prevalence of microbial mats during this part of the Earth's history, particularly in storm-affected subtidal environments and the intertidal zone \cite{Mata2009, Noffke2010}. The abundance of MISS on Precambrian, particularly Proterozoic, siliciclastic bedding planes suggests that microbial mats were widespread at that time \cite{Bose2012}. The absence of metazoan grazing and bioturbation may also have made it  more likely that mats from this period were preserved \cite{Seilacher1999}. Microbially induced structures are also observed in Phanerozoic siliciclastic sediments \cite{Mata2009}, although not as widely as in the Proterozoic geologic record \cite{Porada2007, Samanta2011,Eriksson2012,Parizot2005, Noffke2010}.

The presence and influence of microbial mats can be inferred from sediment properties that are uncharacteristic of sand and mud deposited via a purely physical process \cite{Schieber1999, Cuadrado2012,Dupraz2009, Bailey2006, Bouougri2002}. In ancient siliciclastic biolaminates, former microbial mats are indicated by darker layers rich in iron-oxides, black carbonaceous materials \cite{Bouougri2007, Noffke2010, Noffke2002} and pyrite \cite{Pfluger1996}. These compounds are produced in modern mats by the metabolic activity of micro-organisms living in and below the mat \cite{Dupraz2004,Sarkar2008, Schieber2007a}. In sediments below and above the mat layers, isolated sedimentary grains and mica flakes surrounded by sericitic layers have been attributed to the presence of microbial mats and bacteria, which trapped and bound the grains \cite{Black1933,Noffke2003,Draganits2004}. Isolated and orientated sediment grains can indicate biofilms, which initially formed around the grains and have subsequently grown into full mats, rotating and trapping grains in the process \cite{Noffke1997,Noffke2011}.  Other microbial mat signatures include the presence of tunnel patterns by undermat miners \cite{Crimes1982, Gingras2011}, structures related to mat growth and destruction processes, such as shrinkage cracks \cite{Schieber1999} and wavy laminae. In some cases fossilised microbial filaments \cite{Noffke2003b,Callow2009}, biomat fragments \cite{Steiner2001} and microbial death masks can also act as indicators \cite{Darroch2012}.

\subsection{Physical properties of modern microbial mats}

The presence of microbial mats has been preserved in the ancient geologic record. However, deriving or deducing their exact material properties from fossil findings is not possible. This makes clear-cut statements about the behaviour and properties of ancient microbial mats problematic.  Modern microbial mats are dense, cohesive organic layers that act as a single unit. They are formed of microbes held together by extracellular polymeric substances (EPS) \cite{Wingender1999}, which participate in the binding of cells and the formation of microbial aggregates. EPS act as a cohesive gel-like network, provide a scaffold for the cells and make up 50-90\% of the total organic material in the film \cite{Flemming2010}. EPS have a number of functions \cite{Flemming2010}. They are responsible for adhesion of the mat to surfaces \cite{Donlan2002} and provide mechanical stability for the microbial colony \cite{Mayer1999}. EPS can also stabilise clastic sediment surfaces \cite{Schieber2007}.

The composition, chemical and physical properties of EPS, and therefore microbial colonies, can vary widely. Recent investigations of modern microbial mats however, have revealed some well conserved features. Modern microbial mats behave like viscoelastic films \cite{Vinogradov2004, Rupp2005, Shaw2004, Lieleg2011};  they undergo both reversible elastic responses and irreversible deformation. For example, microbial colonies grown for long periods under turbulent flow conditions have been known to display a rippled surface texture \cite{Stoodley1999}.  Lieleg \emph{et al.} \cite{Lieleg2011} have shown that biofilms display elastic-like responses for high frequency stimuli and viscous-fluid responses when low frequency stimuli are applied. In the context of microbial mats, this means that for short term exposure of the mat to shear stress the mat responds elastically; when the applied stress is removed the mat returns to its original shape. For sustained exposure to shear stress, on the other hand, internal physical stresses are dissipated through viscous flow. This reduces the possibility of structural failure and uncontrolled detachment of the mat under shear.  The shear modulus \emph{G} and viscosity $\eta$ of microbial mats are seen to vary over seven orders of magnitude \cite{Shaw2004}. However, the stress relaxation time $\tau = \eta/G$  is strikingly well conserved across a wide range of biofilms and found to be approximately 18\,min. The timescale $\tau$ is essentially the time over which a biofilm, deformed by external forces, will `forget its original shape'.

From these observations two general conclusions about the properties of microbial mats can be made \cite{Shaw2004}. Firstly, microbial mats behave as viscoelastic films, with little or no influence from biological processes on short time scales. This suggests that the physical properties are solely determined by the EPS. Secondly, the relaxation time  of the viscoelastic medium is universally about 18\,mins, the typical lifetime of the temporary crosslinks in the EPS.

Comparison between modern and ancient microbial structures suggest that ancient microorganisms existed with the same diversity as today \cite{Noffke2010}. It is not unreasonable therefore that ancient and modern microbial mats shared similar general material characteristics. For the purpose of this paper it is assumed that ancient microbial mats can be modelled as viscoelastic polymeric materials with viscosities and relaxation times similar to those of modern microbial mats. The properties of the polymer films are given in section III A.

\subsection{Kinneyia}

Wrinkle and ripple-like structures are perhaps the most documented microbially mediated sedimentary structures \cite{Porada2007, Mata2009,Pruss2004,Parizot2005, Noffke2001}. Observed wrinkle-structures have widely varying morphologies, characteristics, formation processes and preservation modes. Patterns can develop in positive or negative relief \cite{Banerjee2005} and can occur on or within the rock bed \cite{Noffke2001}. There are many different hypotheses as to how wrinkle structures formed, but all agree that a cohesive mat is present at the sediment surface. This is confirmed by the observation of a large variety of wrinkle structures in modern microbial environments \cite{Noffke2010,Gerdes1993, Noffke1999,Noffke1998}.

\begin{figure}
\includegraphics[width=13.5cm]{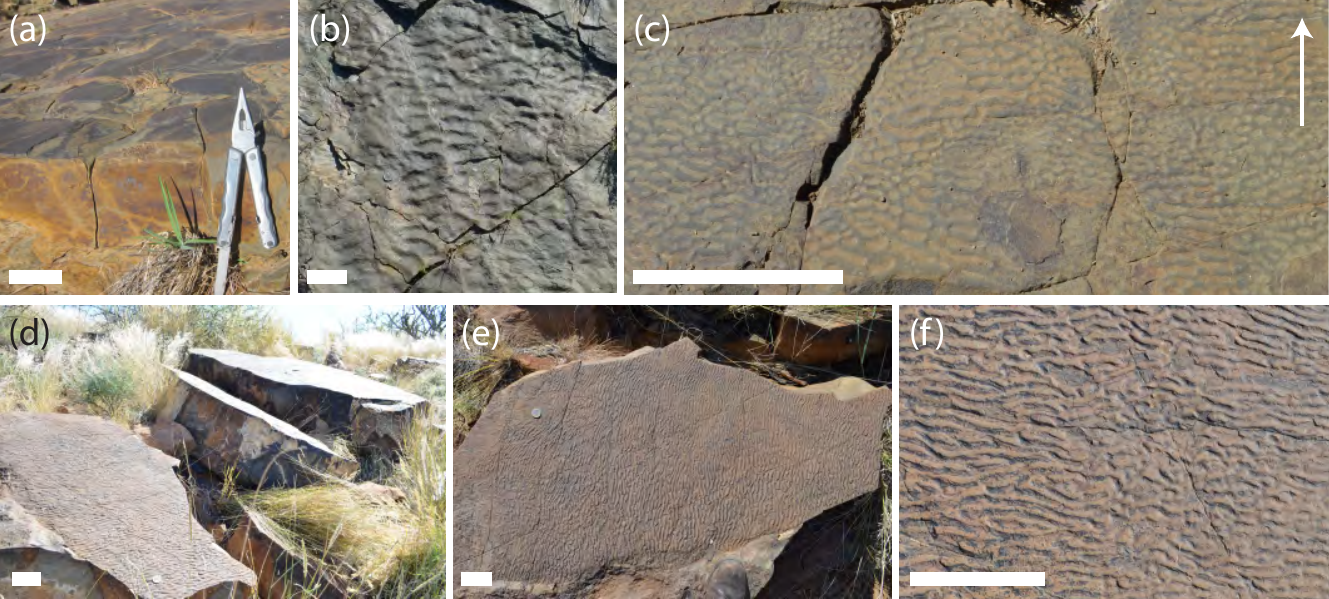}
\caption{\label{Fig:Kinneyia} Proterozoic Kinneyia structures from (a-c) Haruchas farm and (d-f) Neuras farm, Namibia. (a, d) Storm deposits on which Kinneyia are observed. (b) Ripple structures found below the storm deposit. These have a much larger wavelength than the Kinneyia and a different morphology. (c) Example Kinneyia from Haruchas. Arrow shows inclination direction of the outcrop. (e) At Neuras Kinneyia were observed over areas larger than 1\,m$^2$. (f) Close up of (e) showing a clearly defined structure. Scale bars all 10\,cm.}
\end{figure}

The focus of this paper is Kinneyia. To avoid confusion with other types of wrinkle structures, here only Kinneyia as defined by Porada \emph{et al.} \cite{Porada2008} are considered. Kinneyia are a sub-class of sedimentary fossils with clearly defined ripple features (e.g. figure~\ref{Fig:Kinneyia}). Kinneyia are dominantly linear structures found on the upper bedding planes of sandstone or siltstone layers. These bedding layers are much thicker than the amplitude of the Kinneyia pattern and are interpreted as storm or flood event deposits (figure~\ref{Fig:Kinneyia}a). The upper layer of the event deposit is generally observed to be covered with a thin topset veneer thought to have sedimented after the event deposit at high flow velocity and shallow water depths. Thin sections made from Kinneyia from Sweden \cite{Porada2008} show that the depth of the veneer is proportional to the amplitude of the ripple structure, with the bottom of the veneer layer coinciding with the troughs of the Kinneyia.

The Kinneyia pattern is characterised by an undulating ripple-like structure (figure~\ref{Fig:Kinneyia}c) with a wavelength on the millimetre to centimetre scale (2-20\,mm). The ripple shape deviates slightly from a sinusoidal wave, with flattened crests and rounded troughs most commonly described in the literature.  However, rounded and even sawtooth-shaped crests are also seen \cite{Porada2008}. The relative widths of the troughs and the ridges varies widely from sample to sample \cite{Pfluger1999}. The ripple-pattern is generally well ordered with the crests orientated parallel to one another. Honeycomb-like patterns with round or elongated pits are also seen with the two morphologies often coexisting on different areas of the same sample \cite{Porada2007, Hagadorn1999}. The Kinneyia patterns often have one preferred crest orientation. In Kinneyia from \"Oland, this orientation was observed to run perpendicular to the paleocurrents from which the event layers were deposited \cite{Martinsson1965}. However, in some examples a second orientation direction is observed at $40-50^{\circ}$ to the first \cite{Porada2008}. This led Porada \emph{et al.} to suggest that some interference phenomena is involved in their formation \cite{Porada2008}.

\subsection{Suggested mechanisms for Kinneyia formation}

Initially, it was assumed that an abiotic mechanism was responsible for the ripple-like patterns observed in Kinneyia. Proposed mechanisms range from perturbation of cohesive sediments in shallow water by wind \cite{Singh1978}, to erosion of sediments by waves \cite{Shrock1948} and foam induced patterns \cite{Allen1966}. However, none of these mechanisms satisfactorily explain the observed Kinneyia ripple patterns. Hagadorn and Bottjer \cite{Hagadorn1997, Hagadorn1999} were the first to suggest that the formation of wrinkle-structures and Kinneyia may in some way involve microbial mats. This assumption was made due to the observed decrease in grain density near the surface of the wrinkled sedimentary fossils, which suggested that sediments were bound at the surface by an organic matrix. The ripples of Kinneyia are often seen to have very steep sides, which it is suggested would be unstable without some 'glue' or microbial mat to bind the grains together once a critical slope is reached. Hagadorn and Bottjer's  conclusion was also made due to the discovery of similar wrinkle patterns in modern microbial mats found growing in the Great Basin, USA \cite{Hagadorn1999}. The modern patterns had a wavelength of a few millimetres with both rounded and pointed ridges, which extend for tens of centimetres, being observed.

Pfl\"uger went on to propose a microbially mediated gas bubble model for the formation of Kinneyia \cite{Pfluger1999}. He suggested that gas rising from a sedimentary substrate can be trapped by a microbial mat, and collect to form bubbles.  Using a mixture of water-saturated sand and sodium bicarbonate, Pfl\"uger was able to directly observe a process whereby the bubbles destabilize the sediment  leaving trace patterns.   Gas domes, polydisperse round shapes up to tens of centimetres in diameter, are observed in modern and ancient microbial mats \cite{Gerdes2007}.  However, gas domes are no longer thought to be responsible for the formation of Kinneyia, as their patterns do not correspond well with the well-defined elongated ridge and troughs structures of Kinneyia.

The most recent model for the development of Kinneyia is that of Porada \emph{et al.}, who proposed that the pattern forms in liquefied sediment confined beneath  microbial mats. Like Pfl\"uger \cite{Pfluger1999}, Porada \emph{et al.} assumed that the microbial mats acts as a barrier to gas and groundwater trapped in the underlying sediment. Here however, the bacteria are assumed to form dispersed colonies in the sediments, where the mats have adhered and grown downwards. The gas then accumulates in pore spaces between the colonies, leading to a local anisotropy in the water saturated sediment on the microscale. Cyclic stressing, from oscillatory water flow, of the overlying microbial layer, causes the underlying sediments to liquefy. This is due to an oscillating pore pressure induced by the change in water depth with tide cycling. The liquefied sand layer is assumed to be several centimetres thick with the overlying mat being 3\,cm thick. The oscillatory pressure changes induce ripple structures in both the liquefied sand and microbial mat. The wavelength of the ripples is on the order of 1\,m for the sand and a few centimetres for the mat. The ripple pattern in the mat induces further local variations in the pore pressure causing seepage and grain lifting. If the liquefied sediment layer becomes very thin, then replication of the overlying small wavelength microbial ripples can occur in this underlying liquefied layer. This condition is satisfied periodically at the seaward boundary of the mat during each tidal cycle. This model requires a very specific number of criteria to be met, along with significant grain lifting, for the formation of Kinneyia on the scale of a few millimetres to centimetres to occur. When these conditions are not met, ripple patterns on the metre scale are instead predicted.

In the model of Porada \emph{et al.} the small-scale ripple patterns are induced first in the overlying mat and then transferred into the underlying sediment. Observation of modern microbial mats, however, suggest that it is not always possible to make a clear distinction between these two layers. Mats are often seen to develop from biofilms which initially form around individual sediment grains and subsequently grow together into a thick layer. The sediments are thus an integral part of the microbial mat. This is particularly the case for mat growth after storm events.

Here we propose a simple model for the formation of Kinneyia ripple patterns, where the key mechanism is a Kelvin-Helmholtz instability (KHI) induced in the mat under flowing water. A KHI naturally gives rise to an undulating structure on the lengthscales typical of Kinneyia. Evidence from analogue experiments is presented and compared with detailed measurements of fossilised Kinneyia. It should be noted that our model does not contradict that of Porada \emph{et al.}, but is in fact indirectly implied, although not discussed.

\section{Formation of Kinneyia via a Kelvin-Helmholtz type instability}

\begin{figure}
\includegraphics[width=13.5cm]{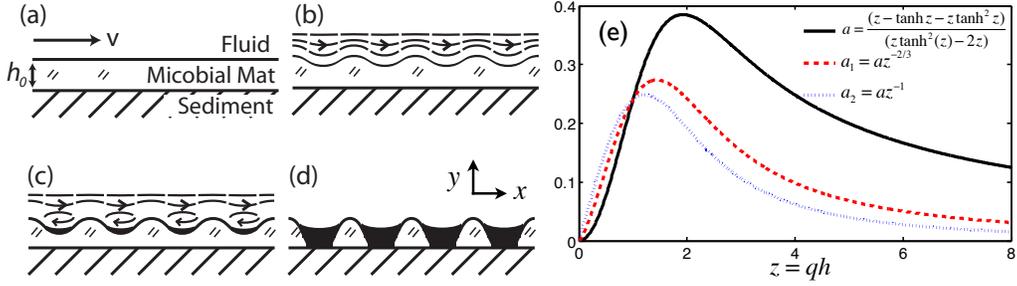}
\caption{\label{Fig:Flow} Schematic representation of the generation of Kinneyia from a hydrodynamic instability. The microbial mat grows in a quiescent environment and is suddenly subject to significant flow in the overlying water (a). This results in a Kelvin-Helmholtz instability, giving rise to a ripple pattern at the surface of the mat (b). Once the amplitude of the ripple reaches a certain threshold, eddies will from in the valleys giving rise to enhanced sedimentation (c). Rupture of the film can occur when the amplitude of the troughs become comparable to the film thickness (d). The predicted relative growth rate (e) of the ripple pattern is sharply peaked around a dominant wavelength where $qh \simeq 2$.}
\end{figure}

Kinneyia are generally found on upper bedding planes in littoral environments that have experienced recent storm deposits \cite{Porada2008}. For the purposes of this model we shall therefore consider a planar microbial mat, or biofilm, on a solid substrate subject to some flow in the overlying fluid (figure~\ref{Fig:Flow}a). The microbial mat is considered to behave as a viscoelastic fluid \cite{Shaw2004}. The system can be approximated by two immiscible fluids (water and microbial mat), with different viscosities and flowing at different velocities over a rigid substrate. The question is then: what happens at the interface between these two fluids?

It is well known that spontaneous destabilisation of a fluid-fluid interface may occur in a two fluid system where the layers respond differently to shear. A well defined instability forms giving rise to a harmonic interfacial corrugation (figure~\ref{Fig:Flow}b). This KHI occurs ubiquitously in nature \cite{Casanova2011, Hasegawa2004, Smyth2012} for example in cloud layers \cite{Dalin2010}. Typically, a KHI is studied using fluids of different densities, however, differences in viscosity also lead to instability.

The system being studied is sketched in figure~\ref{Fig:Flow}. Water flows with far-field velocity $V$ in the $x$-direction over a viscoelastic film of thickness $h$. A flat film/water interface is a solution of the pertinent hydrodynamical equations, but an unstable one. Consider a small periodic perturbation at the interface between the two layers.  Qualitatively the stream lines are compressed at the peaks and expanded at the valleys of the perturbation (figure~\ref{Fig:Flow}b). Due to mass conservation, this requires an increased flow velocity at the peaks and a decreased velocity at the valleys. According to Bernoulli's law this gives rise to a decrease in pressure at the crests and an increase in pressure in the valleys. The pressure differences drive flow in the film from troughs to peaks. Hence small thickness variations in the microbial mat can be amplified over time. For a small perturbation of the interface $h(x,t) = h_0 + \varepsilon (x,t)$, where $\varepsilon (x,t) = \varepsilon_0 (t) \cos qx$ and $q$ is the wave number of the perturbation. The amplitude $\varepsilon_0$ is assumed to vary slowly, such that  the fluid dynamics can be treated quasi-statically.  The velocity, $\mathbf{v} = (v_x,v_y)$, within the viscous layer of the mat is given by Stokes' equation
\begin{equation}
\Delta \mathbf{v} = \frac{1}{\eta}\nabla p
\label{Eq:Stokes}
\end{equation}
where $\eta$ is the viscosity of the microbial mat and $p(x,y,t)$ the pressure within it. The elastic response is neglected here, since only long time dynamics are being considered. The film is taken to be incompressible, such that $\nabla \cdot \mathbf{v} = 0$ everywhere.  From this, it follows that $\Delta p = 0$.  The base of the mat $y=0$ has no-slip boundary conditions, such that $\mathbf{v}(x,0,t)  = 0$, while the upper boundary is taken to be free, such that $\partial_yv_x(x,h,t) = 0$. 

Assuming separation of variables, the general solution to Laplace's equation for the pressure in the mat is
\begin{equation}
p = \varepsilon f(y)= \varepsilon_0(t)\cos{qx} (P_1\cosh{qy} + P_2 \sinh{qy}),
\label{Eq:Pressure}
\end{equation}
where $P_1$ and $P_2$ are constants, which will be determined.  The components of the velocity field that are consistent with the boundary conditions are
\begin{equation}
\mathbf{v} = \big( \varepsilon_0(t) U(y) \sin{qx}, \ \varepsilon_0(t) V(y) \cos{qx}\big)
\end{equation}
for some functions $U(y)$, $V(y)$.  For an incompressible fluid the velocity field can be expressed by derivatives of a scalar stream function $\psi$ as $\mathbf{v} = \nabla \times \psi$.  From this, and Eqn. \ref{Eq:Stokes}, it is derived that $U = -\partial_y V/q$ and $\Delta^2\psi = 0$.  The solution to this latter biharmonic equation yields
\begin{equation}
V = (A_1+B_1y) \cosh{qy} + (A_2 + B_2y)\sinh{qy},
\label{Eq:Stream}
\end{equation}
where $A_1$, $A_2$, $B_1$ and $B_2$ are constants.  The no-slip lower boundary condition $V(0)$ = $\partial_yV(0)$ = 0 implies that $A_1 = 0$ and $A_2 = -B_1/q$.  Combining Eqns. \ref{Eq:Stokes} -- \ref{Eq:Stream} the constants $B_1$ and $B_2$ are found to be $B_1 = P_1/2\eta$ and $B_2 = P_2/2\eta$.  The free boundary condition at $y=h$ gives
\begin{equation}
\partial_{yy}V(h) = B_1(q\sinh{qh}+q^2h\cosh{qh})  + B_2(2q\cosh{qh}+q^2h\sinh{qh}) = 0.
\end{equation}
Thus
\begin{equation}
B_2 = -B_1\bigg(\frac{qh+\tanh{qh}}{2+qh\tanh{qh}}\bigg) = -g(qh)B_1.
\label{Eq:B1}
\end{equation}

If the pressure at the interface is now taken to be
\begin{equation}
p(x,h,t) = -P(q)\varepsilon (x,t)
\label{Eq:PressureVariation}
\end{equation}
then, using Eqns. \ref{Eq:Pressure} and \ref{Eq:B1},
\begin{equation}
B_1 = \frac{P(q)}{2\eta(g(qh)\sinh{qh}-\cosh{qh})}.
\end{equation}
Introducing this back into Eqn. \ref{Eq:Stream} gives 
\begin{equation}
V = \frac{P(q)}{2\eta q} \left[\frac{qy\cosh{qy}-(1+g(qh)qy)\sinh{qy}}{g(qh)\sinh{qh}-\cosh{qh}}\right].
\end{equation}

The equation of motion for the interface is $\partial_t\varepsilon = v_y(x,h,t)$. Thus, $\partial_t\varepsilon_0 = \varepsilon_0V(h)$ and $\varepsilon_0 (t) \propto \exp (\alpha t)$. The growth rate $\alpha(q)$ is
\begin{equation}
\alpha (q) = \frac{P(q)h}{\eta}\left[\frac{qh - \tanh{qh}-qh\tanh^2{qh}}{qh(\tanh^2{qh}-2)}\right].
\label{Eq:Alpha}
\end{equation}
The most rapidly growing mode is given  by the maximum of $\alpha (q)$. Since the expression in brackets is sharply peaked around $qh\approx 2$, the maximum will not be strongly dependent on the exact form of $P(q)$. 

To be more specific, however, an estimate for $P(q)$ needs to be found.  Calculating the pressure distribution above the perturbation exactly, requires the  full boundary layer theory to be considered. A treatment of this problem was carried out by Bordner \cite{Bordner1978}. He found that the pressure scales as $\varepsilon/\delta$, where $\delta$ is the thickness of the disturbance sublayer in the fluid flowing above the mat. The expression for $\delta$ is given by 
\begin{equation}
\delta = \left(\frac{\mu^2 q^2}{4\pi^2 \rho \tau}\right)^{1/3}
\label{Eq:DisturbanceSublayer}
\end{equation}
where $\tau$ is the mean surface shear stress at the boundary.  $\mu$ and $\rho$ are the viscosity and density of the flowing liquid respectively. $P(q)$ is then given by
\begin{equation}
P(q) \propto \left(\frac{4\pi^2 \rho \tau h^2}{\mu^2}\right)^{1/3} (qh)^{-2/3}.
\label{Eq:PressureResult}
\end{equation}
The maximum of $\alpha$ is then $qh\approx 1.4$. Sediment deposited from the flowing water will accumulate in the troughs of the corrugation. The corresponding pressure contribution is then proportional to the wavelength, such that $P(q)\propto 1/q$. This would yield a maximum of $\alpha$ at $qh \approx 1.2$. The expected value of $qh$ should lie between these values, and depend on the relative strength of the contributions. These growth rate predictions are shown in figure~\ref{Fig:Flow}e. In terms of the wavelength of the most unstable mode, an instability of wavelength approximately 4--5 times the mat thickness $h$ is predicted.  

A Kelvin-Helmholtz type instability of a microbial mat therefore leads robustly to the formation of a ripple instability under any shear flow. The most unstable wavelength of the instability is predicted to be proportional to, and a few times, the film thickness, but is insensitive to either the fluid flow speed or mat viscosity. This complies well with the observation that Kinneyia patches often exhibit reduced wavelengths at their boundaries, where the mat can be assumed to have been thinner.  Once the pressure variation is firmly established viscous flow within the film leads to an increasing corrugation amplitude. When the amplitude of the ripple reaches a certain threshold, eddies will form in the valleys of the undulation (figure~\ref{Fig:Flow}c). Some of the clastic sediments carried by the overlying fluid as it flows, will settle in the troughs due to the back flow and stagnation points arising from the eddies. Accumulation of sediments in the troughs may also act as an additional driving force for the instability; the sediment is denser than the microbial mat and the surrounding water. The growth of the ripple pattern is thereby accelerated and steep slopes develop between the troughs and the crests.  As the microbial mat dies, remains of the EPS glue the sediments together, preserving the Kinneyia structures that are found in the geologic record.

\section{Experimental procedure}

To test the hypothesis that Kinneyia arise from KHIs in microbial mats, analogue experiments were carried out where the microbial mat was replaced with an abiotic polymeric viscoelastic film and subjected to flowing water.

\subsection{Materials and methods}

Experiments were carried out using a viscoelastic cross-linked poly(vinyl alcohol) (PVA) film. The PVA films were made by fully dissolving PVA (Sigma-Aldrich, molecular weight 145\,kg/mol) in de-ionized (Millipore) water at $90^{\circ}$C. The mixture was cooled and crosslinked using sodium borate solution (Borax, Sigma-Aldrich). The ratio of PVA to sodium borate was kept constant at 10:1 w/w. White paint (titania-based) was added to the PVA (2\,g/100\,ml) prior to crosslinking to create opaque films. PVA solutions of 3, 4 and 5\% by weight were mixed. The rheological properties of the films were characterised, at $T=20\pm2^{\circ}$C, using a parallel-plate rheometer (Stresstech, Rheosystems). Viscosities were measured in the linear regime for applied stresses of 0.01-100\,Pa. The viscosities of the cross-linked solutions were measured to be $25 \pm 4$\,Pa\,s, $124 \pm 10$\,Pa\,s and $398 \pm 20$\,Pa\,s for 3, 4 and 5\% solutions respectively. The relaxation time of the PVA solutions were found to be 15-18\,s, 115-338\,s and 1200-1450\,s respectively. The variation arises from small differences in PVA and sodium borate concentrations between subsequent batches of the solutions. A relaxation time of 1200-1450\,s corresponds well with the properties of modern microbial mats, which are universally observed to have a relaxation time of around 1020\,s \cite{Shaw2004}. The viscosities of modern mats are seen to vary from 10-10$^9$\,Pa\,s.

\begin{figure}
\includegraphics[width=13.5cm]{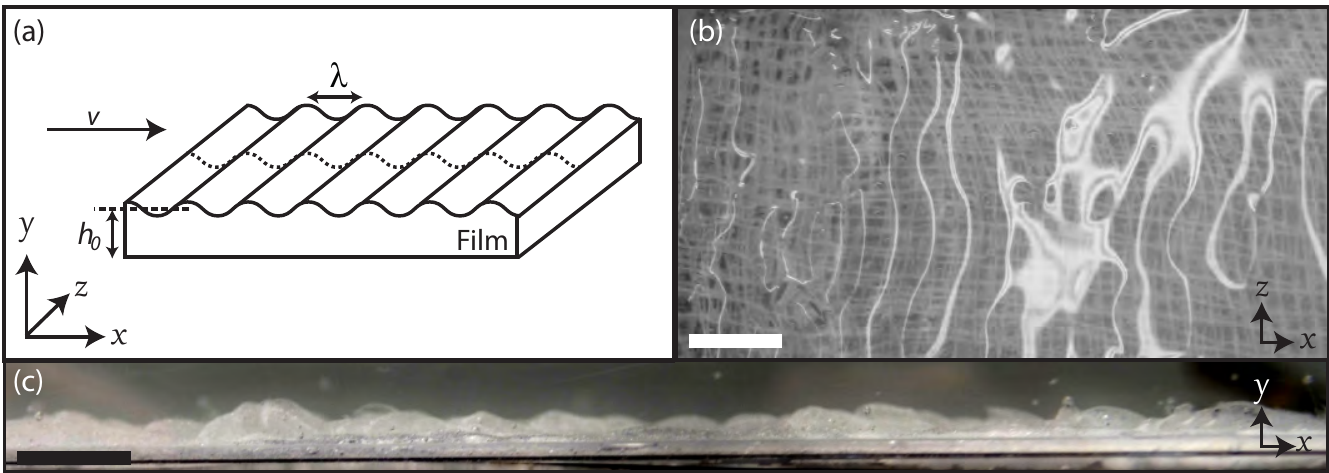}
\caption{\label{Fig:Setup} (a) Deformed PVA film subject to water flow. $h_0$ is the initial thickness of the film and $\lambda$ is the wavelength of the instability. The dashed line represents the deviation of the laser sheet due to the deformed film. (b) Top view of instabilities indicating an elongated sinusoidal pattern.  (c) Side view of instabilities observed in the PVA after 30\,mins of water flow. A sinusoidal pattern is observed. Scale bars in (b,c) are 10\,mm.}
\end{figure}

Two different flow setups were built to probe the behaviour of the films. A schematic diagram of a deformed PVA film subjected to water flow can be seen in figure~\ref{Fig:Setup}a.  The first setup consisted of a small flow cell (9\,$\times$\,6\,$\times$\,2\,cm,\emph{ x}\,$\times$\,\emph{z}\,$\times$\,\emph{y}) connected to a fluid reservoir via a centrifugal pump. The PVA film was placed onto a porous glass substrate (Robu Glas) in the flow cell and left for 1\,hr to relax.  The height of the substrate could be altered, allowing the thickness $h_0$ of the film to be varied. A valve directly in front of the pump allowed the flow speed to be adjusted. Flow speeds of 0.05, 0.12, 0.18 and 0.24\,m/s were used. Glass windows in the flow cell allowed optical observation of the ripple formation. The profile of the film was monitored \emph{in situ} using an oblique 532\,nm laser sheet and imaged every 60\,s using a Nikon D5100 digital SLR camera  mounted directly above the flow cell. The position of the laser line was extracted from the resulting images with a resolution of $\pm0.04$\,mm in the \emph{x}-direction and $\pm0.02$\,mm in the \emph{z}-direction. The wavelength was calculated from the first non-zero peak of the 1D autocorrelation function, the inverse fourier transform of the power spectrum, of the extracted laser line.  The temperature of the reservoir varied from $18-22^{\circ}$C between experiments, but was kept at $\pm 0.5^{\circ}$C during each experiment using a Julabo FT402 temperature controller. All flow experiments were carried out at room temperature ($T=20\pm2^{\circ}$C).

The second setup was a larger flow trough, allowing for a bigger sample (20\,$\times$\,20\,cm) to be observed. This setup was used to qualitatively observe how the patterns developed when sediments were added into the system. The sediments used were glass beads with a diameter of $0.09-0.15$\,mm. Again the height of the substrate could be altered to change the thickness of the film. The upper water surface was in this case free. The thickness of the flowing water layer was $2-3$\,cm depending on the flow speed. Experiments were run at flow speeds of 0.024, 0.085 and 0.17\,m/s. Sediment was deposited uniformly (0.05\,g/cm) over the PVA surface 60\,s after the water flow had started. A camera directly above the sample was used to monitor the development of the instabilities.

\subsection{Fossil measurements}

For comparison with the experimental data, Kinneyia were studied from two sites in Namibia. Both sites date to the terminal Proterozoic Vingerbreek Member, Schwarzrand Subgroup, Nama Group. The first site was located at Haruchas farm, Namibia [24$^{\circ}$ 21' 46.3'' S; 16$^{\circ}$ 24' 21.6'' E] \cite{Noffke2002}. The outcrop is approximately 3\,$\times$\,6\,m in area and is covered with small Kinneyia patches ranging in size from a few cm$^2$ to around 400\,cm$^2$ (e.g. figure~\ref{Fig:Kinneyia}c). The Kinneyia cover 50-60\% of the outcrop. The outcrop sits in a modern dry river bed on top of an event deposit $15-20$\,cm thick (figure~\ref{Fig:Kinneyia}a). This deposit is thought to have arisen from a storm event \cite{Porada2008} and overlies an older rippled substrate, which indicates the flow direction. The older ripples run parallel to the Kinneyia ripples, but have a larger wavelength (10-20\,cm) and a smoother sinusoidal undulation (see figure~\ref{Fig:Kinneyia}b).  Examples of Kinneyia from this specific locality at Haruchas have previously been presented in the literature \cite{Porada2008}.

The second fossil site was located at Neuras farm, Namibia [24$^{\circ}$ 24' 11.3'' S; 16$^{\circ}$ 15' 8.7'' E]. The Kinneyia here are found on two isolated rock outcrops located on a small cliff overlooking a river bed. In this case the event deposit on which the Kinneyia were observed was $\sim$35\,cm thick. The structures observed at Neuras were more extensive than those at Haruchas and were seen to cover areas in excess of 1\,m$^2$. The source of the Kinneyia at Neuras could not be directly observed. However, both Neuras and Haruchas are located along the same storm-affected Proterozoic shoreline.

Replicas of the fossils were made to allow detailed measurements to be made without destroying or removing the outcrops. The areas were cleaned and then cast with Mold Star\textsuperscript{\textregistered}  pourable silicone rubber (Smooth-on). The rubber was left to cure for 45\,min before being peeled back to create a negative replica mold of the underlying structure. Solid positive replicas were produced from these molds using Smooth-Cast\textsuperscript{\textregistered} 300 (Smooth-on) in the laboratory.

Surface profile measurements of the Kinneyia were carried out using a stylus profilometer (Dektak XT Bruker). Areas 400-2500\,mm$^2$ were scanned to create 3D height maps. The 3D map was created by stitching together a series of 2D line traces taken at 50\,\textmu m intervals. The resolution in the line traces was 3\,\textmu m in the horizontal direction and 10\,nm in the vertical direction. The 2D auto-correlation functions of the topographs were computed using Matlab. An angle-dependent auto-correlation function was found by averaging over any azimuthal angle $\theta\pm0.5^{\circ}$. The orientation of the pattern was defined as the value of $\theta$ at which the radial auto-correlation function was at a maximum and the first non-trivial peak of the radial auto-correlation function as the wavelength of the Kinneyia ripples. The peak-to-trough amplitude of the ripples was measured directly from the line scans.

\section{Results and Discussion}
\subsection{Lab-made Kinneyia}

\begin{figure}
\includegraphics{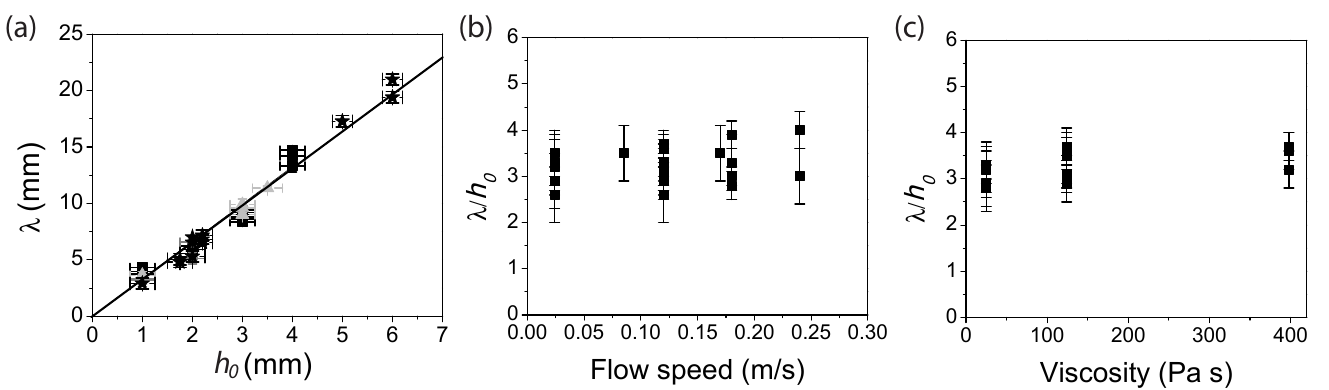}
\caption{\label{Fig:Exp} Saturation wavelength of the ripple instability induced in PVA films subject to flow.  The wavelength does not vary with the flow speed or film viscosity. (a)  All data. PVA ($\blacksquare$) $\eta=124$\,Pa\,s,  ($\medbullet$) $\eta=25$\,Pa\,s and ($\blacktriangle$) $\eta=398$\,Pa\,s at ($\blacksquare$) 0.12\,m/s, (\textcolor[gray]{0.5}{$\blacksquare$}) 0.18\,m/s and (\textcolor[gray]{0.8}{$\blacksquare$}) 0.24\,m/s.  Data were measured using the flow cell. ($\bigstar$) The wavelength of the patterns which develop when sediments are added to the system. Data were measured using the flow trough. The line shows a fit to the data for $\lambda=3.3\pm0.3h_0$, where $h_0$ is the thickness of the initial PVA film. The flow rate and viscosity have no effect on the resulting dominant wavelength that develops.  (b)  $\lambda/h_0$ as a function of flow speed for PVA $\eta=124$\,Pa\,s. (c) $\lambda/h_0$ as a function of viscosity for a flow speed of 0.12\,m/s.  }
\end{figure}

\begin{figure}
\includegraphics{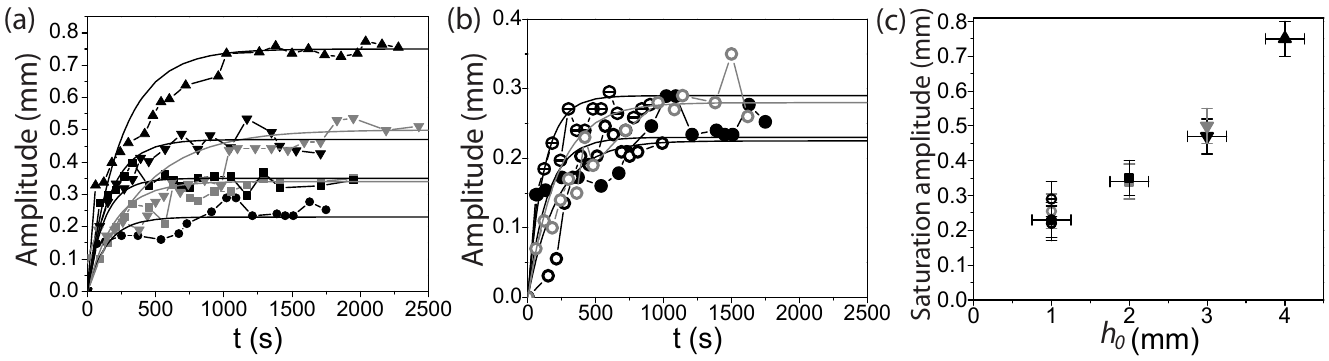}
\caption{\label{Fig:Exp2} Time evolution of the peak to trough amplitude of ripples  induced in PVA films subject to flow. (a) Amplitude as a function of film thickness for ($\medbullet$) 1mm, ($\blacksquare$) 2mm, ($\blacktriangledown$) 3mm and ($\blacktriangle$) 4mm thick films. Grey and black symbols show different runs using the same experimental conditions. Lines show the exponential fits to the data with relaxation times of between 145 and 400\,s. This corresponds to the measured relaxation time of the PVA of 115-338\,s. The variation here is seen for different batches of the PVA mixture. (b) The saturation amplitude is not dependent on the flow rate of the water.  The graph shows data for 1\,mm thick films at flow rates of ($\medbullet$) 0.12\,m/s, ($\bigcirc$) 0.18\,m/s and ($\ominus$) 0.24\,m\,s$^{-1}$. The saturation amplitude is only dependent on the film thickness (c).}
\end{figure}

Ripple patterns are observed to form at the film/water interface in flat PVA films subjected to water flow conditions (figure~\ref{Fig:Setup}b,c). These instabilities are visible within tens of seconds of flow initiation. The growth rate of the instabilities depends on the viscosity of the film. Figure~\ref{Fig:Exp}a shows the wavelength of the instability that forms as a function of the film thickness $h_0$. The wavelength $\lambda$ was monitored using a laser sheet (figure~\ref{Fig:Setup}a) and was found from the first non-zero peak of the 1D autocorrelation function of the deflected laser line. The data points in figure~\ref{Fig:Exp}a are an average of the wavelengths calculated from successive images taken every 60\,s over a 20-45\,minute period. The error bars indicate the standard deviation in the wavelength measured from these images over the course of the experiment. For films with initial thicknesses $h_0$ of 1 to 4\,mm, the wavelength of the ripple pattern that forms is $\sim$3 to 14\,mm, respectively. The line in figure~\ref{Fig:Exp}a shows a linear least-squares fit to the data with a confidence range of $\lambda = 3.3\pm0.3 h_0$. No dependency of the wavelength on either the viscosity of the film or the flow rate is observed (figure~\ref{Fig:Exp}b,c). This is in agreement with the theory presented here, which predicts that the wavelength is only dependent on the thickness of the film.

The growth of the interfacial ripples as a function of time is indicated in figure~\ref{Fig:Exp2}a. The peak-to-trough amplitude of the ripples was measured directly from the  laser line. The amplitude grows with time, eventually saturating. The amplitude saturates on the same timescale as the relaxation time of the polymer. The saturation timescale was not observed to be dependent on the film thickness or the flow rate (figure~\ref{Fig:Exp2}a,b). The saturation amplitude varies linearly with the film thickness (figure~\ref{Fig:Exp2}c). A small increase in the saturation amplitude is observed with increasing flow rate (figure~\ref{Fig:Exp2}b). However, this increase is within the experimental error. The data shown in figure~\ref{Fig:Exp2} were obtained by following the growth of individual ripples along the instability. While the amplitude should not be taken as an absolute value for all ripples formed under the given conditions, they are representative and show the trends observed in the experimental data.

\begin{figure}
\includegraphics[width = 13.5cm]{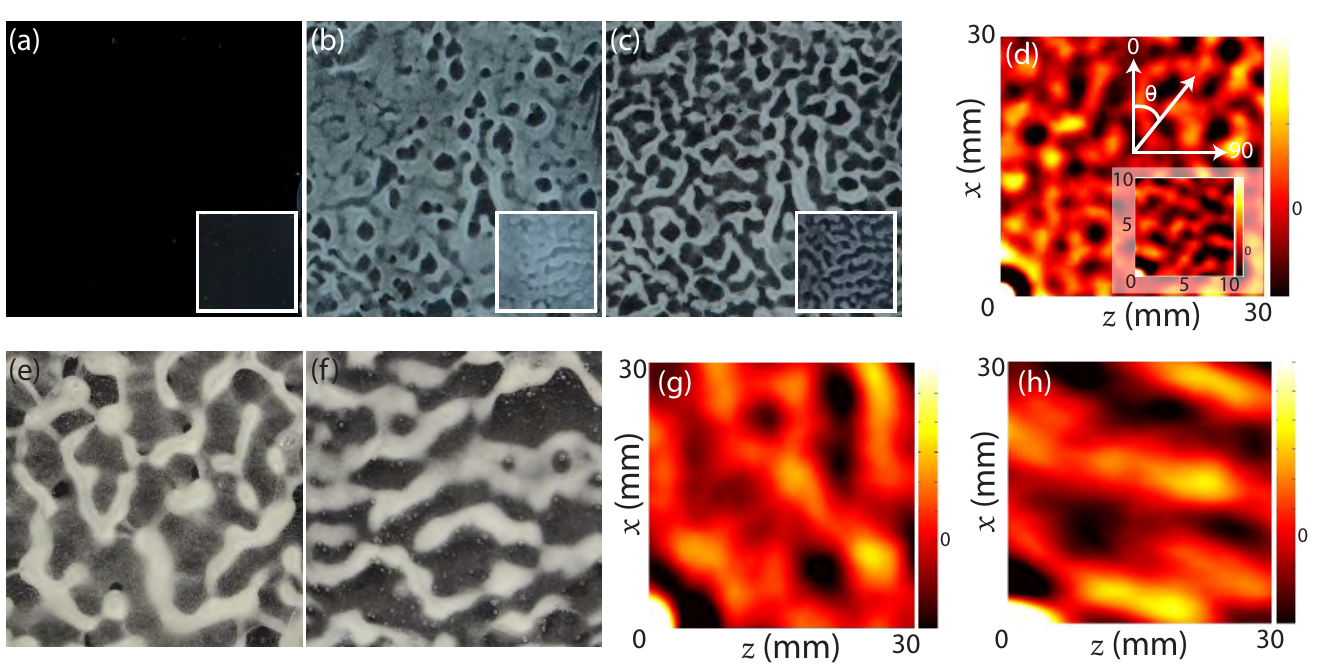}
\caption{\label{Fig:ExpKin} Development of lab-Kinneyia in a 2\,mm thick PVA film at a flow speed of 0.024\,m/s. Prior to flow the film is initially flat (a). The black colour comes from the underlying substrate. After 60\,s of flow sediment is deposited. Images (b) and (c) show the pattern formation 120\,s and 180\,s after the flow started.  All images are 6\,$\times$\,6\,cm. The insets (2\,$\times$\,2\,cm) show the patterns at the edge of the film where it was thinner (1\,mm thick). A decrease in the wavelength is observed in the thinner film.   (d) The 2D autocorrelation function of (c). The first non-zero peak is observed at 0$^{\circ}$ at a wavelength of 5.2\,mm, where $0^{\circ}$ is defined to be parallel to the flow direction.  The inset in (d) shows the 2D autocorrelation function of the inset in (c). The first non-zero peak is observed at 0$^{\circ}$ for a wavelength of 2.9\,mm. (e-h) Development of lab-Kinneyia in 6\,mm thick films for flow speeds of (e) 0.024\,m/s and (f) 0.18\,m/s. (g, h) 2D autocorrelation function of (e, f). The wavelengths are 19.4\,mm and 21\,mm in (e) and (f) respectively. An increase in the order is observed with increasing flow speed. (e, f) are the same size as (a-c). The colour maps for the 2D autocorrelation functions show the intensity in arbitrary units.}
\end{figure}

In our model of  Kinneyia formation eddies in the ripple valleys lead to enhanced sedimentation (figure~\ref{Fig:Flow}c). To test the effects of sedimentation in our  PVA films, glass beads were added to the system while the water was flowing. The glass beads used were large enough for sedimentation to occur. As the instabilities developed, the beads collected in the troughs of the pattern (figure~\ref{Fig:ExpKin}). The resulting, experimentally produced, PVA Kinneyia (hereafter lab-Kinneyia) that form are well-ordered and exhibit an interconnected pattern of elongated ridges. The same thickness-wavelength relationship observed in the other experiments (figure~\ref{Fig:Exp}a) is found. Overspill of the PVA, from the sample chamber, leads to the development of patterns in regions with two different thicknesses on the same sample. The patterns that develop have the same morphology, but different wavelengths (inset figure~\ref{Fig:ExpKin}b,c). This shows that Kinneyia patterns with different wavelengths can arise under the same flow conditions due to thickness variations across the sample.

Figure~\ref{Fig:ExpKin}e, f shows the lab-Kinneyia that formed in 6\,mm thick films at low and high flow speeds. The pattern in figure~\ref{Fig:ExpKin}e is qualitatively similar, but with a longer wavelength, to that shown in figure~\ref{Fig:ExpKin}c for the 2\,mm thick film. The patterns in figure~\ref{Fig:ExpKin}c and \ref{Fig:ExpKin}e were formed at the same flow speed. Increasing the flow speed appears to result in an increase in the order of the pattern, with a transition from honeycomb-like patterns to parallel ridges perpendicular to the flow direction. The increase in order is observed in the 2D autocorrelation functions of the resulting patterns (figure~\ref{Fig:ExpKin}g,h). The patterns generated in the lab-Kinneyia shown in figure~\ref{Fig:ExpKin}  qualitatively resemble the Kinneyia found in Namibia e.g. figures~\ref{Fig:Kinneyia} and \ref{Fig:FT}.  The results suggest that the different patterns may arise from variations in the flow conditions under which the microbial mat or film is deformed.

\subsection{Kinneyia}

\begin{figure}
\includegraphics[width = 13.5cm]{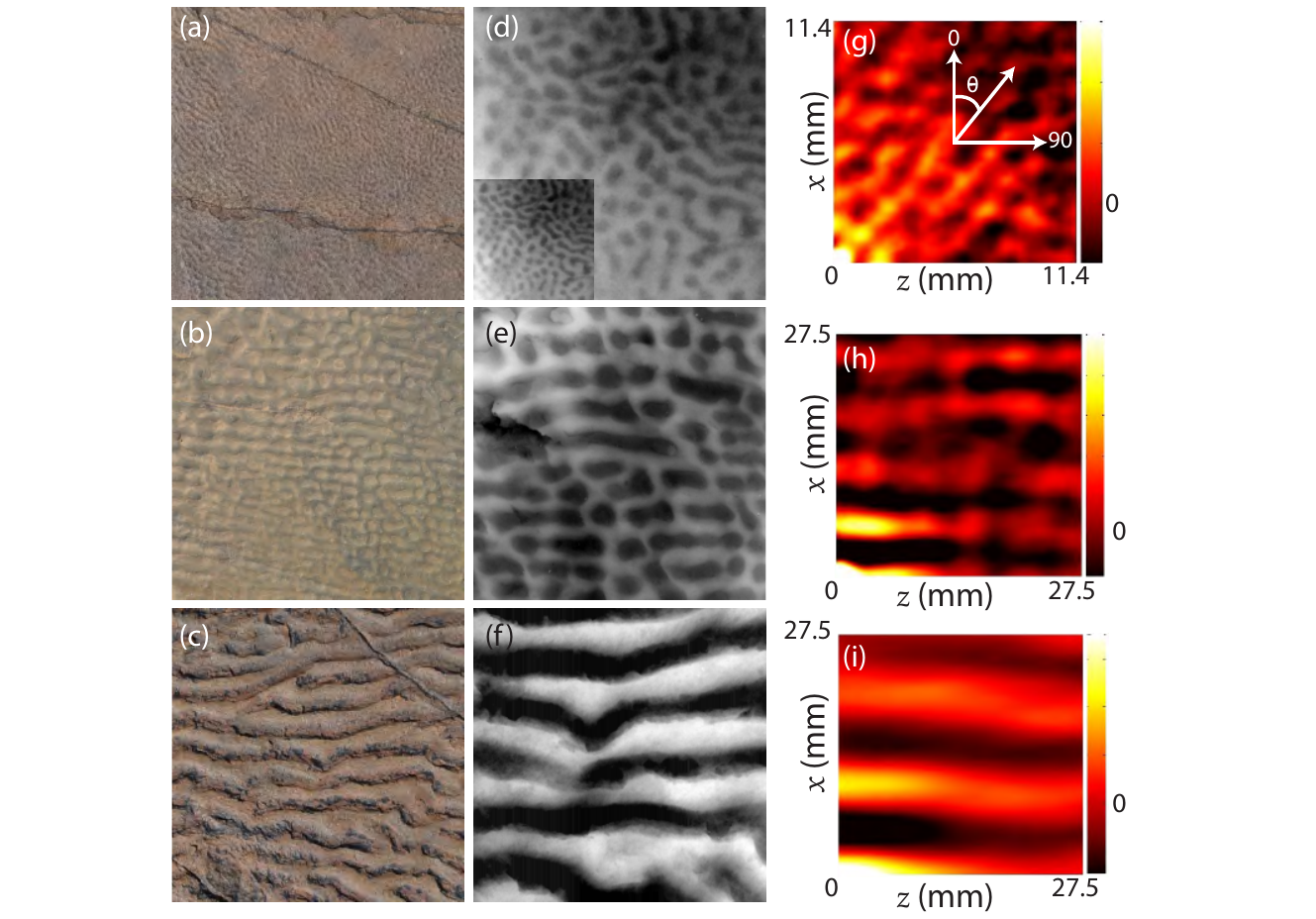}
\caption{\label{Fig:FT} Photos of Kinneyia stuctures found at (a, c) Neuras farm and (b) Haruchas farm. The photographs in (a-c) have the same magnification and are 10\,x\,10\,cm. (d-f) Three-dimensional map scans taken from casts of (a-c) respectively. The scans are (d) 22.5$\times$22.5\,mm and (e, f) 55$\times$55\,mm. The inset in (d) shows the map scan in (d) at the same magnification as (e) and (f). (g-i) Two-dimensional autocorrelation functions of (d-f). The colour maps show the relative intensity of the peaks in arbitrary units. The arrows in (g) indicate the orientation of the ripples. }
\end{figure}

Figure~\ref{Fig:FT} shows representative height profiles of the Kinneyia replicas made at Neuras and Haruchas. In each case images of the fossil (figure~\ref{Fig:FT}a-c) and map scan (figure~\ref{Fig:FT}d-f)  are shown alongside the 2D autocorrelation function of the map scan (figure~\ref{Fig:FT}g-i) . Morphologies range from small circular-shaped patterns (figure~\ref{Fig:FT}a) to elongated ridges extending across the whole fossil (figure~\ref{Fig:FT}c). Intermediate patterns containing circular and elongated ridges are also seen (figure~\ref{Fig:FT}b).

\begin{figure}
\includegraphics[width=13.5cm]{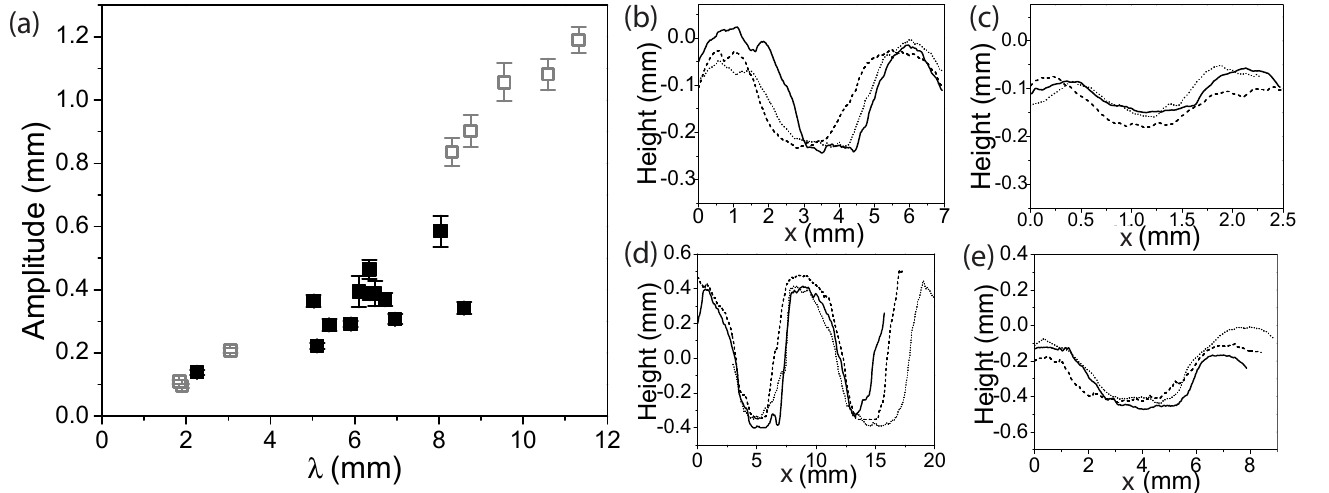}
\caption{\label{Fig:Dektak} (a) Amplitude of the Kinneyia ripples as a function of their wavelength for fossils from ($\blacksquare$) Haruchas ($\Box$) Neuras. (b-e) Example line traces for fossils from (b), (d) Haruchas and (c), (e) Neuras. Line traces are for wavelengths from $\sim$1-10mm. The ripples have a flattened sinusoidal morphology.}
\end{figure}

Analysis of the profilometry measurements from the Kinneyia can be seen in figure~\ref{Fig:Dektak}. The points in figure~\ref{Fig:Dektak}a give the wavelength and corresponding amplitude for each area scanned. The data show that as the wavelength of the ripple increases so to does the amplitude. This is in agreement with the trends observed in the analogue PVA flow experiments (figures~\ref{Fig:Exp} and \ref{Fig:Exp2}). The amplitude measured from the fossils however, can only be taken as a lower limit for the amplitude of the instability when it first formed. Compaction after burial may have considerably decreased the amplitude of the ripples. In addition, the outcrops at Haruchas and Neuras are unprotected from the elements and subject to erosion. It is therefore likely that the original amplitude was higher. It is not possible however, to quantify the effects of either compaction or erosion. The amplitude of the Kinneyia ripples (figure~\ref{Fig:Dektak}a) are higher than those for the PVA experiments (figure~\ref{Fig:Exp2}). The PVA experiments from figure~\ref{Fig:Exp2}  do not take into account the effect of sedimentation, which was qualitatively observed to enhance the growth and break up of the ripples (figure~\ref{Fig:Flow}d). Comparison of figure~\ref{Fig:Dektak}a and figure~\ref{Fig:Exp}a suggests that the microbial mats involved in Kinneyia formation were $\sim 0.5-4$\, mm thick. This thickness range corresponds well with the  literature values for modern mats \cite{Neu1997,Gerdes2000}. While the original thickness of the ancient microbial mats involved in Kinneyia formation can not be measured directly from the fossils, these measurements suggest that the ancient microbial mats had similar thicknesses.

Example line traces from four Kinneyia samples are shown in figure~\ref{Fig:Dektak}b-e. Line scans for fossils with wavelengths varying from 1.8 to 10.6\,mm are shown. The line scans in figures~\ref{Fig:Dektak}c and \ref{Fig:Dektak}d correspond to the fossils shown in figures~\ref{Fig:FT}a and \ref{Fig:FT}c respectively. It has previously been stated that Kinneyia ripples are characterised by sinusoidal waves with flattened crests and rounded troughs \cite{Porada2008}, although no direct evidence for this has been provided. While in some cases the line scans do indicate flattened crests and rounded troughs (e.g. figure~\ref{Fig:Dektak}e), this morphology is not universal. Troughs and crests can be both ``rounded" and ``flattened". However, in all cases a periodic wave is observed. When considering the shape of the ripples, in particular the shape of the crests, factors that may have altered their shape over time should  be taken into account. It is not unreasonable to assume that rounded crests may become more flattened due to the effects of erosion or compaction. Erosion could presumably also lead to flattened crests becoming more rounded.

The orientation of the ripples was measured from the autocorrelation function of the three-dimensional map scans (figure~\ref{Fig:FT}g-i). For well-ordered ripples peaks are observed at $0^{\circ}$ and $90^{\circ}$ (e.g. figure~\ref{Fig:FT}h,i).  Here 0$^{\circ}$ is defined to be perpendicular to the ripples (\emph{x}-axis), while $90^{\circ}$ runs parallel to the ripples (\emph{z}-axis), as indicated by the arrows in figure~\ref{Fig:FT}g. In well-ordered patterns the peaks at  $0^{\circ}$ and $90^{\circ}$ correspond to the wavelength and mean extent of the ripples respectively. For more disordered ripples (e.g. figure~\ref{Fig:FT}a) additional peaks are observed at other angles $\theta$ in the two-dimensional autocorrelation function.  The angle $\theta$ at which the additional peaks are observed defines an additional orientation direction of the ripples and is indicated by the arrow in figure~\ref{Fig:FT}g. $\theta$ varies slightly from sample to sample, but lies between 30$^{\circ}$ and 50$^{\circ}$ with a mean value of 33$\pm$5$^{\circ}$. A secondary orientation was also noted by Porada \emph{et al.} \cite{Porada2008}. A similar secondary orientation is observed in the lab-made Kinneyia shown in figure~\ref{Fig:ExpKin}.

There are a number of possible origins for this secondary orientation, including multi-directional flows arising from tidal rips or differences in the forward motion of the tide and the subsequent backwash, changes in the flow direction or the presence of obstacles in the flow path. Another possibility is that the secondary orientation is induced by the ripples themselves. Pinning of the upper air/water surface to the PVA ripple crests is observed in the analogue PVA/sediment experiments when the amplitude of the ripple becomes comparable to the height of the flowing water layer. This induces small wakes, at the air/water interface, locally changing the flow direction of the water. Over time this secondary flow could influence the orientation of the ripples that form in the film. However, a full experimental and theoretical study of the 2D stability of Kinneyia ripples is beyond the scope of the current paper.

The robustness of the mechanism put forward here  suggests that Kinneyia type structures should be abundantly found in modern biomats. However, wrinkle patternss are only very occasionally observed in modern biomats and generally have morphologies that differ from Kinneyia patterns. To resolve this apparent contradiction, one should bear in mind that, for some so far unknown reason, Kinneyia fossils are intimately linked to storm deposits, which represent rather scarse events. In fact, Kinneyia is a comparably rare fossil. In addition, the omnipresence of grazing animals and rapid bioturbation since the  Cambrian may have rendered the conditions for Kinneyia formation increasingly unfavorable, to the point of their virtual absence today.

Another, very speculative but nevertheless interesting, line of thought concerns the rheological properties of biomats. Although modern biofilms may appear to be viscoelastic media to a rheometer, we know that active mechano-response is well conserved at least in modern eucaryotes. Strictly speaking, biomats must therefore be viewed as active matter, possibly stalling the instability observed in abiotic films. As we do not know the history of mechano-response evolution, the mechanical properties of biomats may have been, subtly but distinctly, different when Kinneyia fossils formed.

\section{Conclusions}

Destabilisation of viscoelastic films under shear by flowing water results in the formation of sinusoidal ripple-like structures. The key process involved in the ripple formation is a Kelvin-Helmholtz instability, which arises from the spontaneous destabilisation of a fluid-fluid interface due to shear, when the two fluids have different viscosities. The wavelength of the ripples is dependent on the thickness of the film, but not on the film's viscosity or the flow speed. Changes in the flow speed result in changes in the ordering of the ripples, with both honeycomb-like patterns and well-ordered parallel ripples being observed. The wavelength and morphology of the ripples corresponds well with the patterns seen in Kinneyia structues. The experimental results from the lab-Kinneyia suggest that the microbial mats involved in Kinneyia formation were $\sim0.5-4$\,mm thick. This fits well with the thicknesses of modern biofilms and microbial mats found in the literature \cite{Gerdes2000, Neu1997}.

The formation of Kinneyia via a Kelvin-Helmholtz instability accurately replicates the pattern formation process in viscoelastic microbial mats. The littoral areas where Kinneyia are frequently observed would have been perfect habitats for microbial mats to develop and grow. Storm events would have provided strong hydrodynamic flows, deforming the mats in an undulative manner and would also have provided the sediments necessary to preserve this pattern through rapid burial. Both the experiments and theoretical predications here suggest that Kinneyia-like rippling is a universal behaviour, explaining the abundance of Kinneyia in the ancient geologic record.

\begin{acknowledgments}
The authors would like to thank Reinhold Wittig for his help when carrying out the field work in Namibia.
\end{acknowledgments}

\bibliographystyle{plainnat}

\end{document}